\documentclass[aps,prc,twocolumn,superscriptaddress,nofootinbib]{revtex4-1}

\usepackage{arydshln}
\usepackage{multirow}
\usepackage{amsmath}    % need for subequations
\usepackage{mathrsfs}   % fancy F
\usepackage{amssymb}
\usepackage{graphicx}   % need for figures
\usepackage{verbatim}   % useful for program listings
\usepackage{color}      % use if color is used in text
\usepackage{subfigure}  % use for side-by-side figures
\usepackage{hyperref}   % use for hypertext links, including those to external documents and URLs
\usepackage{natbib}     % use for bibliography

\begin{document}

\title{$Q_{\textrm{EC}}$-value determination for $^{21}$Na$\rightarrow^{21}$Ne and $^{23}$Mg$\rightarrow^{23}$Na mirror-nuclei decays using high-precision mass spectrometry with ISOLTRAP at ISOLDE/CERN}

\author{J.$\>$Karthein$\>$}
    \altaffiliation{This article contains data from the PhD thesis of Jonas Karthein, enrolled at the {Ruprecht-Karls-Universit\"at} Heidelberg}
    \email[\\Corresponding author: ]{jonas.karthein@cern.ch}
    \affiliation{CERN, Route de Meyrin, 1211 Gen\`eve, Switzerland}
    \affiliation{Max-Planck-Institut f\"ur Kernphysik, 69117 Heidelberg, Germany}
\author{D.$\>$Atanasov}
    \altaffiliation[Present address: ]{KU Leuven, Instituut voor Kern- en Stralingsfysica, 3001 Leuven, Belgium}
    \affiliation{Max-Planck-Institut f\"ur Kernphysik, 69117 Heidelberg, Germany}
\author{K.$\>$Blaum}
    \affiliation{Max-Planck-Institut f\"ur Kernphysik, 69117 Heidelberg, Germany}
\author{M.$\>$Breitenfeldt}
    \affiliation{CERN, Route de Meyrin, 1211 Gen\`eve, Switzerland}
\author{V.$\>$Bondar}
    \affiliation{KU Leuven, Instituut voor Kern- en Stralingsfysica, 3001 Leuven, Belgium}
\author{S.$\>$George}
    \affiliation{Max-Planck-Institut f\"ur Kernphysik, 69117 Heidelberg, Germany}
\author{L.$\>$Hayen}
    \affiliation{KU Leuven, Instituut voor Kern- en Stralingsfysica, 3001 Leuven, Belgium}
\author{D.$\>$Lunney}
    \affiliation{CSNSM-IN2P3-CNRS, Universit\'e Paris-Sud, 91405 Orsay, France}
\author{V.$\>$Manea}
    \altaffiliation[Present address: ]{KU Leuven, Instituut voor Kern- en Stralingsfysica, 3001 Leuven, Belgium}
    \affiliation{CERN, Route de Meyrin, 1211 Gen\`eve, Switzerland}
\author{M.$\>$Mougeot}
    \altaffiliation[Present address: ]{Max-Planck-Institut f\"ur Kernphysik, 69117 Heidelberg, Germany}
    \affiliation{CSNSM-IN2P3-CNRS, Universit\'e Paris-Sud, 91405 Orsay, France}
\author{D.$\>$Neidherr}
    \affiliation{GSI Helmholtzzentrum f\"ur Schwerionenforschung, 64291 Darmstadt, Germany}
\author{L.$\>$Schweikhard}
    \affiliation{Universit\"at Greifswald, Institut f\"ur Physik, 17487 Greifswald, Germany}
\author{N.$\>$Severijns}
    \affiliation{KU Leuven, Instituut voor Kern- en Stralingsfysica, 3001 Leuven, Belgium}
\author{A.$\>$Welker}
    \affiliation{CERN, Route de Meyrin, 1211 Gen\`eve, Switzerland}
    \affiliation{Technische Universit\"at Dresden, 01069 Dresden, Germany}
\author{F.$\>$Wienholtz}
    \affiliation{CERN, Route de Meyrin, 1211 Gen\`eve, Switzerland}
    \affiliation{Universit\"at Greifswald, Institut f\"ur Physik, 17487 Greifswald, Germany}
\author{R.N.$\>$Wolf}
    \altaffiliation[Present address: ]{ARC Center of Excellence for Engineered Quantum Systems, The University of Sydney, NSW 2006, Australia}
    \affiliation{Max-Planck-Institut f\"ur Kernphysik, 69117 Heidelberg, Germany}
\author{K.$\>$Zuber}
    \affiliation{Technische Universit\"at Dresden, 01069 Dresden, Germany}

\date{\today}

\begin{abstract}
\noindent We report on high-precision $Q_{\textrm{EC}}$ values of the $^{21}$Na$\rightarrow^{21}$Ne and $^{23}$Mg$\rightarrow^{23}$Na mirror $\beta$-transitions from mass measurements with ISOLTRAP at ISOLDE/CERN. A precision of $\delta m/m = 9 \times 10^{-10}$ and $\delta m/m = 1.5 \times 10^{-9}$ was reached for the masses of $^{21}$Na and $^{23}$Mg, respectively. We reduce the uncertainty of the $Q_{\textrm{EC}}$ values by a factor five, making them the most precise experimental input data for the calculation of the corrected $\mathscr{F} t$-value of these mixed Fermi/Gamow-Teller transitions. For the $^{21}$Na$\rightarrow^{21}$Ne $Q_{\textrm{EC}}$ value, a $2.3 \sigma$ deviation from the literature $Q_{\textrm{EC}}$-value was found.

\end{abstract}

\maketitle

% ==============================================================================
% $$$$$$$$$$$$$$$$$$$$$$$$$$$$$$$$$$$$$$$$$$$$$$$$$$$$$$$$$$$$$$$$$$$$$$$$$$$$$$
% ==============================================================================
%                                 INTRODUCTION
% ==============================================================================
% $$$$$$$$$$$$$$$$$$$$$$$$$$$$$$$$$$$$$$$$$$$$$$$$$$$$$$$$$$$$$$$$$$$$$$$$$$$$$$
% ==============================================================================

\section{Introduction}

\noindent After more than five decades of experiments determining half-lives, $Q_{\textrm{EC}}$ values and branching ratios for a set of 14 superallowed Fermi $\beta$-transitions, a very robust data set has been obtained, leading to an impressive $2\times10^{-4}$ precision on the weighted average corrected $\mathscr{F} t$ value for these transitions \cite{Hardy2015}. The constancy of these corrected $\mathscr{F} t$ values confirms the Conserved Vector Current (CVC) hypothesis \cite{Feynman1958} and provides a very precise value for the dominant $V_{\textrm{ud}}$ \textit{up-down} quark-mixing matrix element \cite{Hardy2015}. Together with the $V_{\textrm{us}}$ and $V_{\textrm{ub}}$ matrix elements the unitarity of the Cabibbo-Kobayashi-Maskawa (CKM) quark-mixing matrix \cite{Tanabashi2018} is now confirmed at the $5.5\times10^{-4}$ precision level \cite{Hardy2015} thereby providing strong constraints on several types of new physics beyond the Standard Model \cite{Vos2015, Gonzalez2018, Czarnecki2018}.\\

\noindent The uncertainty on the weighted averaged $\mathscr{F} t$ value for the superallowed Fermi transitions is mainly determined by the theoretical uncertainty on the nucleus-independent radiative correction, $\Delta_R$ \cite{Marciano2006}. Addressing this again to improve its theoretical uncertainty by at least a factor of 2 to 3 would be highly desirable and would allow for major progress in searches for new physics via the CKM-unitarity condition.\footnote{\label{footnote1} During the preparation of this manuscript a new transition independent correction value was published \cite{Seng2018}. However, since this value has shifted significantly from previous values and since additionally it is breaking the CKM unitarity, we decided to use the one from Ref.$\>$\cite{Severijns2008}.}\\

\noindent Meanwhile, progress from the experimental side is continuously ongoing. Input data for the $\mathscr{F} t$ values of well-known superallowed Fermi transitions are being cross-checked and further improved. In addition, with production means at radioactive beam facilities steadily improving, the set of transitions of interest is being extended as well \cite{Hardy2015}. Finally, it would be of interest to obtain a precise value of $V_{\textrm{ud}}$ from further types of $\beta$-transitions. This would not only allow cross-checking the validity of small theoretical corrections but, if sufficiently precise, would also contribute to further reducing the uncertainty of the $V_{\textrm{ud}}$ value.\\

\noindent The $\beta$-decay of the free neutron requires no nuclear structure-related corrections and would thus in part provide an independent check on the value of $V_{\textrm{ud}}$. This requires the determination of the neutron lifetime and of the ratio of the axial-vector to vector coupling constants $g_A/g_V$. Significant progress in the determination of the neutron life-time has been made over the last decade \cite{Gonzalez2018, Czarnecki2018}. The ratio $g_A/g_V$, traditionally extracted from the electron-emission asymmetry parameter, $A$, faces a similar problem \cite{Brown2018}. However, the most recent and also most precise results, obtained from independent measurements, all seem to converge to a common value \cite{Gonzalez2018, Czarnecki2018}.\\

\noindent It was pointed out that also the superallowed mirror $\beta$-transitions in isospin doublets could contribute to further improving the precision on $V_{\textrm{ud}}$ \cite{Naviliat-Cuncic2009}. Moreover, such transitions could provide important cross-checks for the calculation of the isospin impurity correction, $\delta_C$ \cite{Towner2010}. In the past decade, many measurements leading to more precise $\mathscr{F} t$ values for such transitions have been performed. In addition, all theoretical contributions necessary to obtain the corrected $\mathscr{F} t$ values with a precision at the 10$^{-4}$ level - for sufficiently precise experimental input data - have been provided \cite{Severijns2008}. However, similar to the case of the neutron life-time, these mixed Fermi/Gamow-Teller transitions require the determination of the ratio of the axial-vector to vector part in the decay. For the mirror $\beta$-transitions this mixing ratio has traditionally been extracted from the$\beta$-particle emission-asymmetry parameter, $A$, the$\beta$-neutrino correlation coefficient, $a$, and the neutrino-asymmetry parameter, $B$. As reaching high precision in$\beta$-decay correlation measurements is not straightforward, in most cases the precision on the mixing ratio determines the precision of $V_{\textrm{ud}}$ \cite{Naviliat-Cuncic2009, Severijns2013}. The most precise results for $V_{\textrm{ud}}$ from mirror $\beta$-transitions have been obtained for $^{19}$Ne \cite{Calaprice1975} and $^{37}$K \cite{Fenker2018}. The weighted average of the transitions for which data are available, i.e. $V_{\textrm{ud}}$ = 0.9730(14) \cite{Gonzalez2018}, is still about 7 times less precise than the value from the superallowed Fermi transitions \cite{Hardy2015}.\\

\noindent With the advancement of recent radioactive ion beam facilities, intense $^{21}$Na and $^{23}$Mg beams of high purity are now relatively easy to obtain. Hence the mirror $\beta$-transitions of these two nuclei are ideal cases to further improve the value of $V_{\textrm{ud}}$ from mirror $\beta$-transitions. When the proposal for the experiments reported here was submitted to the ISOLDE and Neutron Time-of-Flight Committee (INTC) at CERN \cite{Breitenfeldt:1551494} the $Q_{\textrm{EC}}$ value of both isotopes were the second-largest fractional contribution to their $\mathscr{F} t$ values. New measurements were reported since by TITAN \cite{Schultz2014} and LEBIT \cite{Eibach2015}. The data presented in this work constitutes the most precise results for the $Q_{\textrm{EC}}$ values of these two isotopes to date. From the three experimental input data to the $\mathscr{F} t$ values, the $Q_{\textrm{EC}}$ value now contributes the smallest fraction of the uncertainty for both isotopes, and provides thus strong motivation for improved measurements of the other quantities.

% ==============================================================================
% $$$$$$$$$$$$$$$$$$$$$$$$$$$$$$$$$$$$$$$$$$$$$$$$$$$$$$$$$$$$$$$$$$$$$$$$$$$$$$
% ==============================================================================
%                              EXPERIMENTAL SETUP
% ==============================================================================
% $$$$$$$$$$$$$$$$$$$$$$$$$$$$$$$$$$$$$$$$$$$$$$$$$$$$$$$$$$$$$$$$$$$$$$$$$$$$$$
% ==============================================================================

\section{Experiment and analysis}

% ==========================
% beam production
% ==========================

\noindent The sodium, neon, and magnesium isotopes discussed in the present article were produced at the ISOLDE facility at CERN \cite{Borge2018}. There, a proton beam of up to $2\,\mu$A at an energy of $1.4\,$GeV from CERN's Proton-Synchrotron booster is impinged on a silicon carbide target \cite{Hagebo1992} to produce the desired isotopes. The target was heated up to 2000$\,^{\circ}$C to allow the release of the produced isotopes via thermal diffusion and effusion. In order to enable re-acceleration to $30\,$keV and magnetic mass separation using the general-purpose separator (GPS), the sodium and neon nuclides were ionized using the recently developed versatile arc discharge and laser ion source (VADLIS) \cite{Kirchner1976, MartinezPalenzuela2018} in its electron-impact ionization mode. The magnesium ions were selectively ionized using ISOLDE's resonant ionization laser ion source (RILIS) \cite{Fedosseev2017} while the VADLIS was used in a special surface-ion suppressing mode \cite{MartinezPalenzuela2018, DayGoodacre2016}.\\

% ==========================
% ISOLTRAP + MR-ToF
% ==========================

\begin{figure}[h!]
\centering
    \includegraphics[width=.8\linewidth]{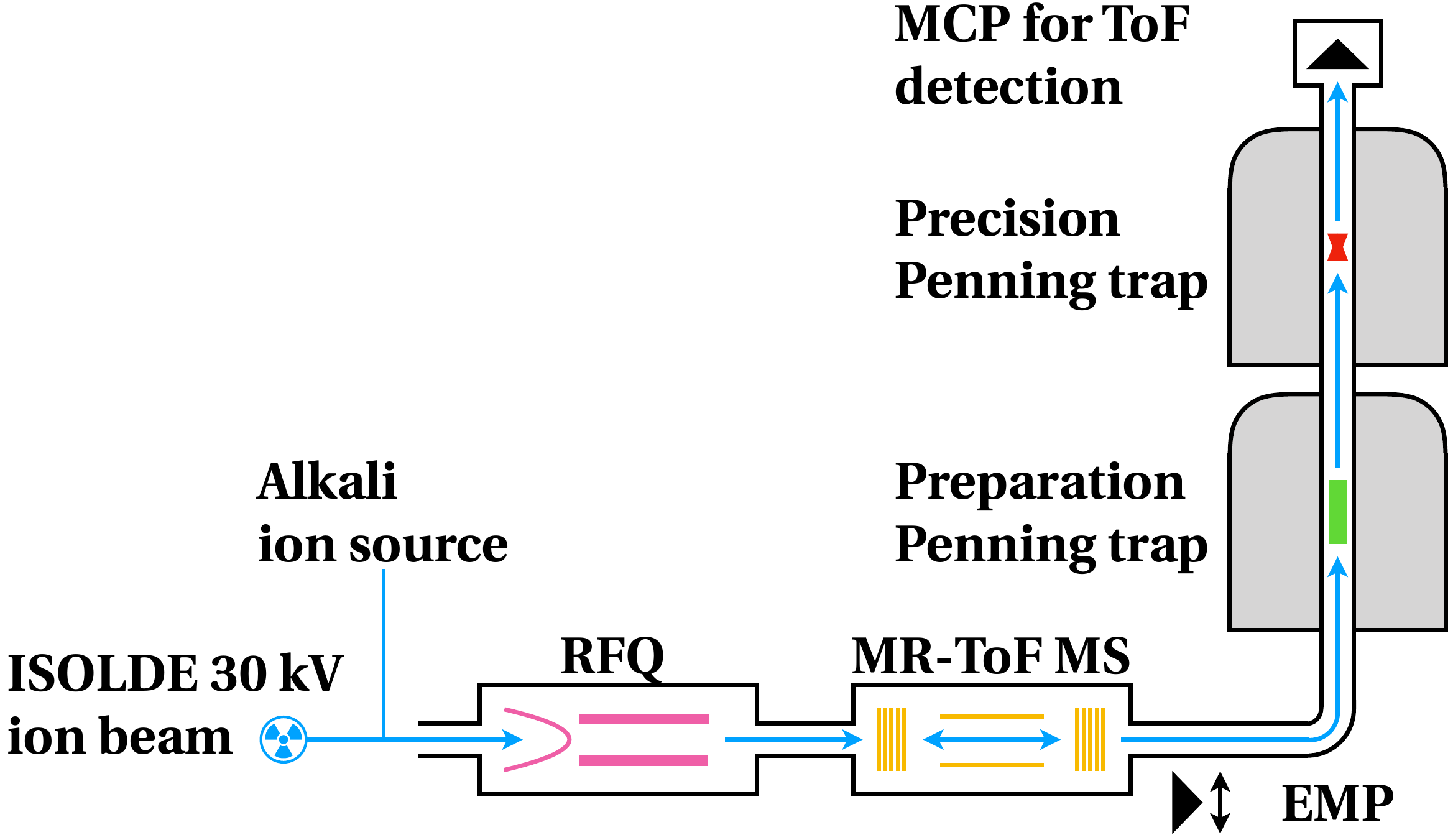}
    \caption{Schematic overview of the ISOLTRAP setup. On-line beam from ISOLDE/CERN or off-line beam from ISOLTRAP's offline source enters to the left to go through a sequence of four ion traps: a linear radio-frequecy Paul trap (RFQ, pink), a multi-reflection time-of-flight (MR-ToF, yellow) device, and two Penning traps (green, red). For particle detection and time-of-flight measurements, a secondary electron multiplier (EMP) ion detector and a micro-channel plate (MCP) ion detector are used.}
    \label{fig:ISOLTRAP}
\end{figure}

\begin{figure}[ht!]
\centering
    \includegraphics[width=\linewidth]{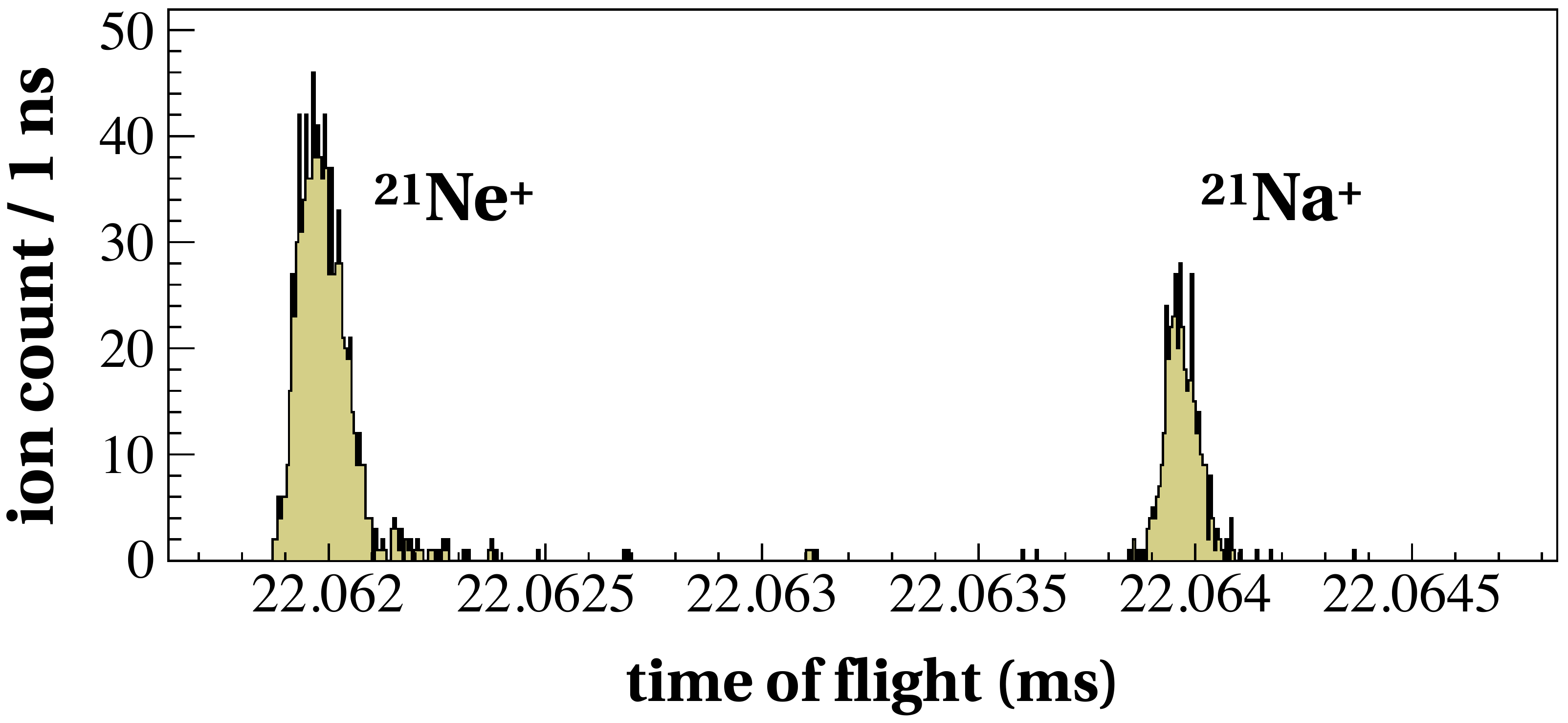}
    \caption{Typical time-of-flight spectrum using laser ionization (for details, see text), here shown for 80 summed spectra of $A=21$ after 2000 revolutions in ISOLTRAP's MR-ToF MS.}
    \label{fig:MR-ToF}
\end{figure}

\noindent The high-precision mass spectrometer ISOLTRAP \cite{Mukherjee2008, Lunney2017, Kreim2013}, schematically depicted in Fig.$\>$\ref{fig:ISOLTRAP}, includes a linear radio-frequency Paul trap (RFQ), a multi-reflection time-of-flight (MR-ToF) device and two Penning traps. The continuous on-line beam from ISOLDE or from the off-line alkali ion source arrives at ISOLTRAP's RFQ (see purple part in Fig.$\>$\ref{fig:ISOLTRAP}) cooler and buncher \cite{Herfurth2001}, which accumulates, bunches and cools the continuous beam in a $1.9\times10^{-3}\,$mbar helium buffer-gas environment for $20\,$ms. The bunched beam is then extracted from the RFQ and its energy is adjusted to $3.2\,$keV using a pulsed drift cavity. The ions are then injected in the MR-ToF mass spectrometer/separator (MS) \cite{Wolf2011, Wolf2013} (see yellow part in Fig.$\>$\ref{fig:ISOLTRAP}). The latter is the first trap which can be used for high-precision mass determination and ion identification. In order to inject/eject ions into/from the MR-ToF MS, a so-called lift cavity situated between the trapping electrodes is switched to ground \cite{Wolf2012, Wolf2012a}. This reduces the kinetic energy of the ions to be lower/higher than the electrostatic trapping potential created by the mirror electrodes. Inside the MR-ToF MS, the ion bunch was reflected between $1000$ and $2000$ times, corresponding to a trapping time of $\sim$ 15 to 25$\,$ms and extending its flight path accordingly. Therefore, ions with the same kinetic energy $E_{\textrm{kin}}=q_iU=m_iv_i^2/2$ (charge $q_i$, acceleration voltage $U$, and velocity $v_i$) and different masses $m_i$ are separated for the same flight path since the mean flight time $t_i$

\begin{equation}
    t_i = \alpha\,\sqrt{m_i/q} + \beta
\end{equation}
is proportional to their mass-over-charge ratio ($\alpha$ and $\beta$ are calibration constants of the ToF-system). After ejection, the ions were detected using a secondary electron multiplier (EMP) ion detector (see Fig.$\>$\ref{fig:MR-ToF}). Once a sufficient time-of-flight separation is achieved, the ion-of-interest (IoI) were selected by properly timing the potential change of the in-trap lift \cite{Wienholtz2017}. In Fig.$\>$\ref{fig:MR-ToF}, the achieved mass resolving power $R$ was on the order of $R=t_i/(2\times\textrm{FWHM}_i)\approx10^5$ (with the mean time of flight $t_i$ and the full width at half maximum FWHM$_i$ of the time-of-flight distribution).\\

% ==========================
% Penning-traps
% ==========================

\noindent The purified beam from the MR-ToF MS then enters the helium buffer-gas-filled preparation Penning trap where the ions are further cooled and purified \cite{Savard1991}. Finally, the IoIs are transferred to the precision Penning trap where the high-precision mass measurements are performed by determining the ions' cyclotron frequency $\nu_c$
\begin{equation}
    \nu_c = \frac{1}{2\pi} \times \frac{q_i}{m_i} B
    \label{eq:cyc.freq}
\end{equation}
with the charge-to-mass ratio $q_i/m_i$ and the magnetic field strength $B$. The detection techniques used in this work were the single-excitation-pulse time-of-flight ion-cyclotron-resonance (ToF-ICR) \cite{Graff1980} and the two-pulse Ramsey-type ToF-ICR \cite{Kretzschmar1999, George2007}. In both cases, a quadrupolar excitation frequency, applied to the trap's segmented ring electrode, is scanned. This couples the two radial eigenmotions of the trapped particles. If the excitation frequency equals the cyclotron frequency of the trapped ion ensemble, their time of flight after ejection to a detector is shorter \cite{Graff1980}.\\

\begin{figure}[ht!]
\centering
    \includegraphics[width=\linewidth]{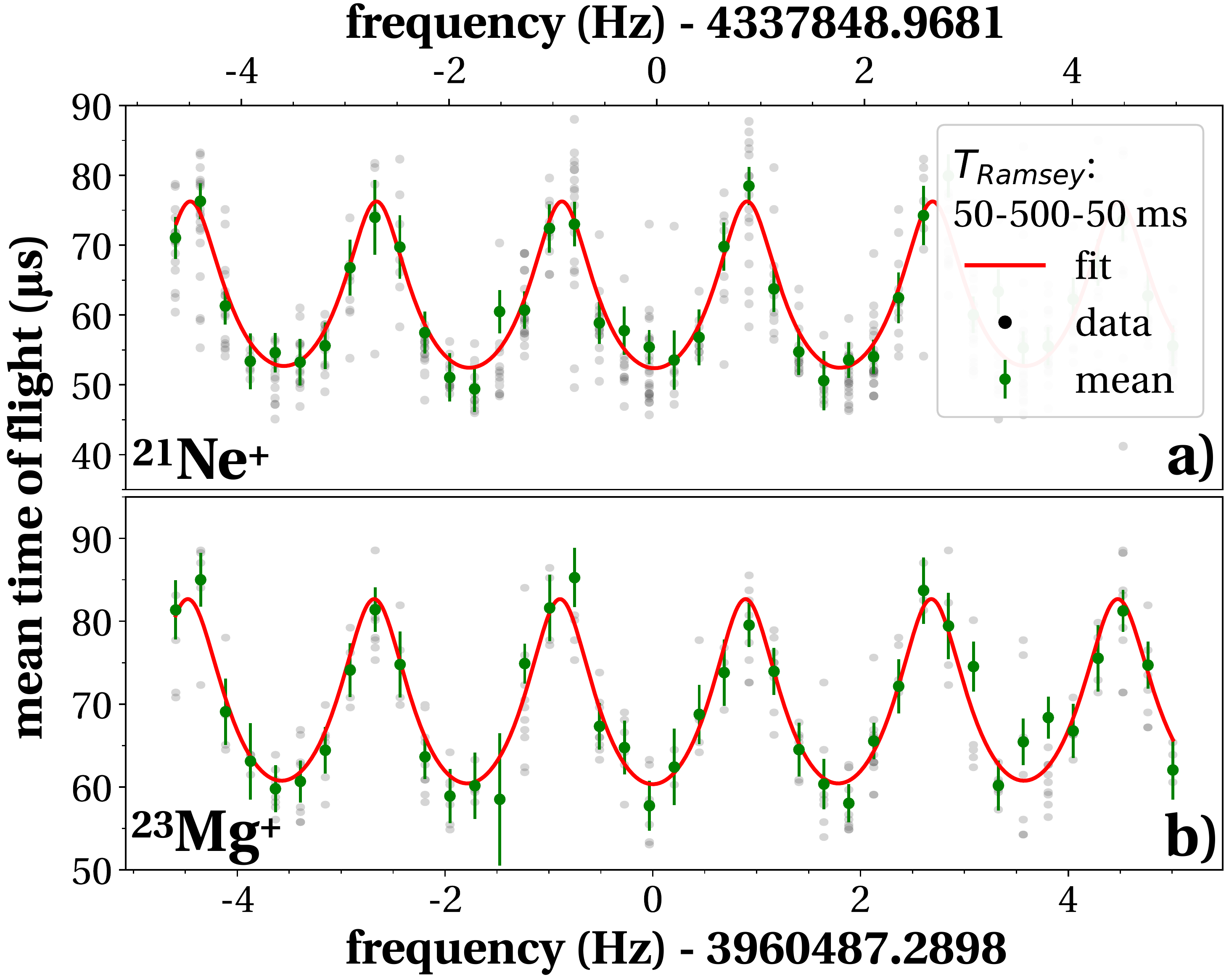}
    \caption{Typical Ramsey-type ion-cyclotron resonances, here shown for $^{21}$Ne$^+$ and for $^{23}$Mg$^+$ after a total of $600\,$ms measurement time (Ramsey pattern: $50\,$ms - $500\,$ms - $50\,$ms) in ISOLTRAP's precision Penning trap. For each frequency bin, the mean of the recorded, unbinned time-of-flight distribution (black) and its associated standard deviation is depicted in green. The red line represents a least squares fit to the expected line shape \cite{Konig1995}.}
    \label{fig:Ramsey}
\end{figure}

\noindent In the case of the $A=21$ system, 30 spectra pairs of subsequent reference/IOI measurements were taken while 19 were taken for $A=23$. In all cases the Ramsey technique (Ramsey pattern: $50\,$ms - $500\,$ms - $50\,$ms, $100\,$ms - $1000\,$ms - $100\,$ms and in case of $^{21}$Ne$^+$ even $200\,$ms - $2000\,$ms - $200\,$ms) was used in order to reduce the statistical uncertainty (see Tab.$\>$\ref{tab:results}). Over the duration of the beam time the mass was switched four times on the ISOLDE mass separator to exclude systematic uncertainties deriving from the data acquisition at ISOLTRAP. Furthermore trap parameters such as the capture time in the trap, the (magnetron) excitation amplitude, the injection voltage, and transport parameters from the preparation trap to the precision trap were varied consistently for both the reference and IOI. With respect to these changes, no statistically significant deviation was observed. Finally, comparison spectra were taken using the well-established single-pulse ToF-ICR technique. A typical Ramsey-type ToF-ICR spectra for $^{21}$Ne$^+$ and $^{23}$Mg$^+$ at an excitation time of $50\,$ms per pulse and a waiting time of $500\,$ms is shown in Fig.$\>$\ref{fig:Ramsey}.\\

% ==========================
% Data
% ==========================

\begin{table*}
\caption{\label{tab:results}Summary for $^{21}$Na$^{+}$ and $^{23}$Mg$^{+}$ showing the number of Ramsey-type spectra taken, the estimated production yield at ISOLDE, the half-lives \cite{Audi2017}, the reference ion for cyclotron frequency ratio determination, the measured cyclotron frequency ratio $r$, and the measured $Q_{\textrm{EC}}$-values using ionization energies from NIST \cite{NIST} in comparison the ones published by LEBIT for $^{21}$Na \cite{Eibach2015} and by TITAN for $^{23}$Mg \cite{Schultz2014}.}
\begin{ruledtabular}
\begin{tabular}{llllllll}
& & & & & &\multicolumn{2}{c}{$Q_{\textrm{EC}}$ (keV)} \\
\cline{7-8}
Isotope  & $\#_{\textrm{spectra}}$ & Yield (s$^{-1}$) & $T_{1/2}$ (s) & Ref.    & Ratio $r$  & This work    & Literature     \\
\hline
$^{21}$Na$^+$  & 30 & $6\times10^6$ &  22.422(10)  & $^{21}$Ne$^+$ & 1.0001813796(9) & 3546.919(18) & 3547.11(9)   \\
\cdashline{1-8}
$^{23}$Mg$^+$  & 19 & $1\times10^8$ &  11.317(11)  & $^{23}$Na$^+$ & 1.0001894144(15) & 4056.179(32) & 4056.35(16)   \\
\end{tabular}
\end{ruledtabular}
\end{table*}

\noindent The time-of-flight spectra were fitted using the well-established analysis software EVA \cite{Kellerbauer2003} while cross-checking additionally selected spectra with a customized analysis software based on ROOT \cite{Antcheva2009}. During the evaluation of the data, the impact of varying the time-of-flight-selection window was systematically investigated. It was found to be well below the statistical uncertainty of the measurement, which can be explained by the purity of the spectra (see Fig.$\>$\ref{fig:MR-ToF}). Therefore, the window was kept the same for all measurements. A $z$-class analysis \cite{Kellerbauer2003}, i.e. an evaluation of the data with respect to the number of ions inside the Penning trap for a given measurement cycle, could not be performed due to the intentionally low count rate of about one ion per cycle. Three independent analyses of the whole dataset following the procedure described in \cite{Kellerbauer2003} were carried out in order to confirm the robustness of the result with respect to the subjective choices made by the evaluators.\\

% ==========================
% Mass calculation
% ==========================

\noindent From alternating cyclotron-frequency measurements between the IoI and the reference nucleus, one can determine the ratio
\begin{equation}
    r=\frac{\nu_{c,\textrm{ref}}}{\nu_{c,\textrm{IoI}}}
    \label{eq:r}
\end{equation}
in order to eliminate systematic uncertainties, e.g. coming from temporal variations of the magnetic field $B$. The well-established calculation procedure uses a linearly interpolated $\nu_{c,\textrm{ref}}$ between the two closest measured cyclotron frequencies of the reference ion at the time of the measurement of the IoI. The final ratio value is then calculated as the weighted mean of all individual ratios. In case of the present measurement series the reference isotope is the daughter nucleus of the corresponding $\beta$-decay and the ion of interest the mother nucleus. This allows direct determination of the $Q_{\textrm{EC}}$ value from the frequency ratio $r$ while minimizing systematic uncertainties:

\begin{equation}
    Q_{\textrm{EC}}=(r-1)\times(m_{\textrm{ref,lit}} - m_e)\, c^2 + r \, E_{i,ref} - E_{i,ioi},
    \label{eq:q_ec}
\end{equation}
with the literature mass for the reference atom $m_{\textrm{ref,lit}}$, the electron mass $m_e$ \cite{Sturm2014}, the speed of light $c$ and the ionization energies for the reference atom and atom of interest, $E_{i,ref}$ and $E_{i,ioi}$ respectively.\\

\begin{figure}[ht!]
\centering
    \includegraphics[width=\linewidth]{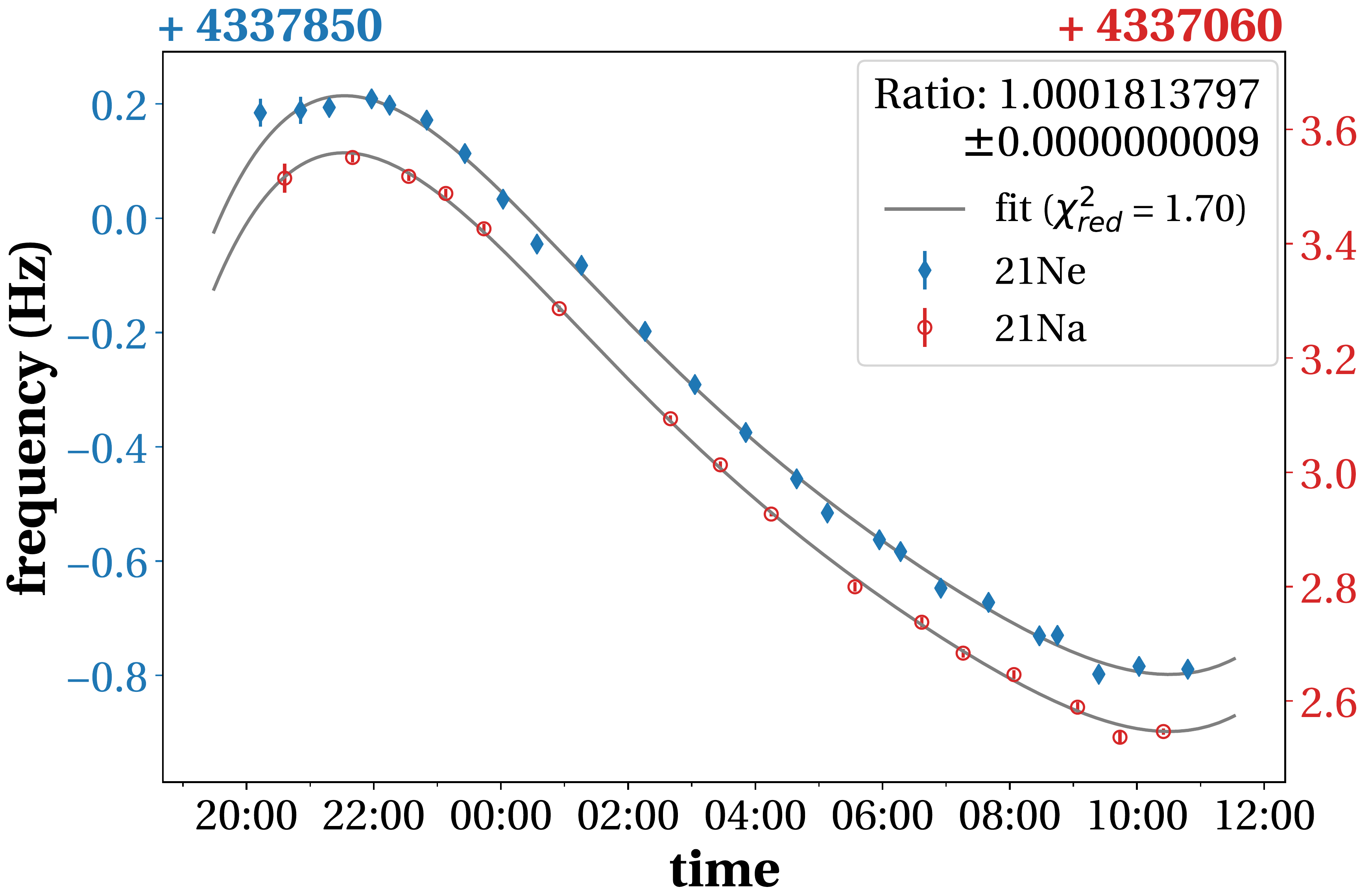}
    \caption{Cyclotron-frequency-ratio determination using a simultaneous polynomial fit for a subset of measured cyclotron frequencies (incl. error bars) of $^{21}$Ne$^+$ and $^{21}$Na$^+$.}
    \label{fig:sim-fit}
\end{figure}

\noindent In addition, the fitting technique described in \cite{Fink2012} which uses a polynomial fit to simultaneously model the temporal evolution of the cyclotron frequency measurements of the mother and daughter nucleus was used. The result of this fit is shown in Fig.$\>$\ref{fig:sim-fit}, where a fifth-order polynomial function was fitted to a subset of measured cyclotron frequencies for both decay partners of mass $A=21$. The proportionality factor between the fit functions represents the cyclotron-frequency ratio for the whole measurement series between the two masses. Therefore, the fluctuations of the measured cyclotron frequencies $\nu_i(t)$ can be described with a polynomial function $f(t)$ and the frequency ratio of Eq.$\>$(\ref{eq:r}):

\begin{eqnarray}
    \nu_{\textrm{IoI}}(t) & = & f(t) \\
    \nu_{\textrm{ref}}(t) & = & r\times\nu_{\textrm{IoI}} = r\times f(t).
\end{eqnarray}

\noindent The degree of the polynomial function describes the dominant effects leading to a change in cyclotron frequency over time. For the presented data it is determined using the degree of the continuously measured magnet's bore temperature fluctuation during the measurement time. The final ratios are calculated as the weighted mean of the fitted subsets. In addition, correlations between fit parameters were calculated and found to be insignificant.\\

\noindent The polynomial fitting technique and the linear interpolation analysis techniques agree well within one combined $\sigma$. Following the description in \cite{Kellerbauer2003}, the mass-dependent effect, the ion production process, the magnetic-field drift, and ISOLTRAP's absolute residual systematic uncertainty were considered. The mass-dependent effect of isobars in this mass range is more than one order of magnitude smaller than the statistical uncertainty and is thus negligible. For each pair, the production process and the experimental conditions were kept constant to avoid any systematic effects. Furthermore the magnetic-field drift is taken into account by the polynomial fitting. The results of the data analysis are summarized in Tab.$\>$\ref{tab:results}. In the case of the  $^{21}$Na$\rightarrow^{21}$Ne transition, a $2.3 \sigma$ deviation from the literature $Q_{\textrm{EC}}$ value, which is dominated by the value reported in \cite{Eibach2015}, was found.

% ==========================
% Calculations table
% ==========================
\newpage

\section{Discussion}

\noindent With the measured $Q_{\textrm{EC}}$-value one can calculate the mirror-nuclei $\mathscr{F} t^{\textrm{mirror}}$-value

\begin{equation}
    \mathscr{F} t^{\textrm{mirror}} = f_V \, t \, (1+\delta^{'}_R) \times (1+\delta^V_{\textrm{NS}}-\delta^V_{\textrm{C}}) \\
\end{equation}

\noindent using the nucleus-dependent radiative corrections, $\delta^{'}_R$ and $\delta^V_{\textrm{NS}}$, and the isospin-symmetry breaking correction $\delta^V_{\textrm{C}}$ calculated in Ref.$\>$\cite{Severijns2008}. The vector parts of the statistical-rate functions $f_V$ were calculated using our new $Q_{\textrm{EC}}$-values and the formalism described in Ref.$\>$\cite{Hayen2018}. The corrected mirror-nuclei $\mathscr{F} t^{\textrm{mirror}}_0$-value can then be calculated according to the relation:

\begin{equation}
    \mathscr{F} t^{\textrm{mirror}}_0 = \mathscr{F} t^{\textrm{mirror}} \times \left( 1 + \frac{f_A}{f_V} \rho^2\right)
\end{equation}

\noindent where $\rho$ is the Fermi/Gamow-Teller mixing ratio while $f_A/f_V$ is the ratio of the axial to vector statistical rate functions. The latter was calculated using the formalism of Ref.$\>$\cite{Hayen2018} and the results of shell model calculations performed with "universal" $sd$ (USDB) interaction (in a full \emph{sd} valence-space) and the$\beta$-spectrum generator (BSG) nuclear shell-model code \cite{Hayen2019}.\\

\noindent For the $^{21}$Na$\rightarrow^{21}$Ne transition the Fermi/Gamow-Teller mixing ratio $\rho=-0.714(7)$ was calculated according to:
\begin{equation}
    \rho = \pm \frac{\sqrt{3-3 \, a_{\textrm{SM}}}}{\sqrt{1+ 3 \, a_{\textrm{SM}}}}
\end{equation}

\noindent using the$\beta$-neutrino asymmetry coefficient $a_{\textrm{SM}}$ from Ref.$\>$\cite{Vetter2008}. The sign of $\rho$ can be derived from the aforementioned shell-model calculations. For $^{23}$Mg there has not yet been a measurement which allows the calculation of $\rho$. The partial half-lives $t$

\begin{equation}
    t = t_{1/2} \times \left(\frac{1+P_{\textrm{EC}}}{BR}\right)
\end{equation}

\noindent were calculated using the half-lives $t_{1/2}$ and branching-ratios $BR$ given in Ref.$\>$\cite{Shidling2018, Grinyer2015, Finlay2017} and in Ref.$\>$\cite{Severijns2008} for $^{21}$Na and $^{23}$Mg, respectively. For both transitions, the correction for the competing electron-capture process $P_{\textrm{EC}}$ was taken from Ref.$\>$\cite{Severijns2008}.\\

\begin{figure}[ht!]
\centering
    \includegraphics[width=\linewidth]{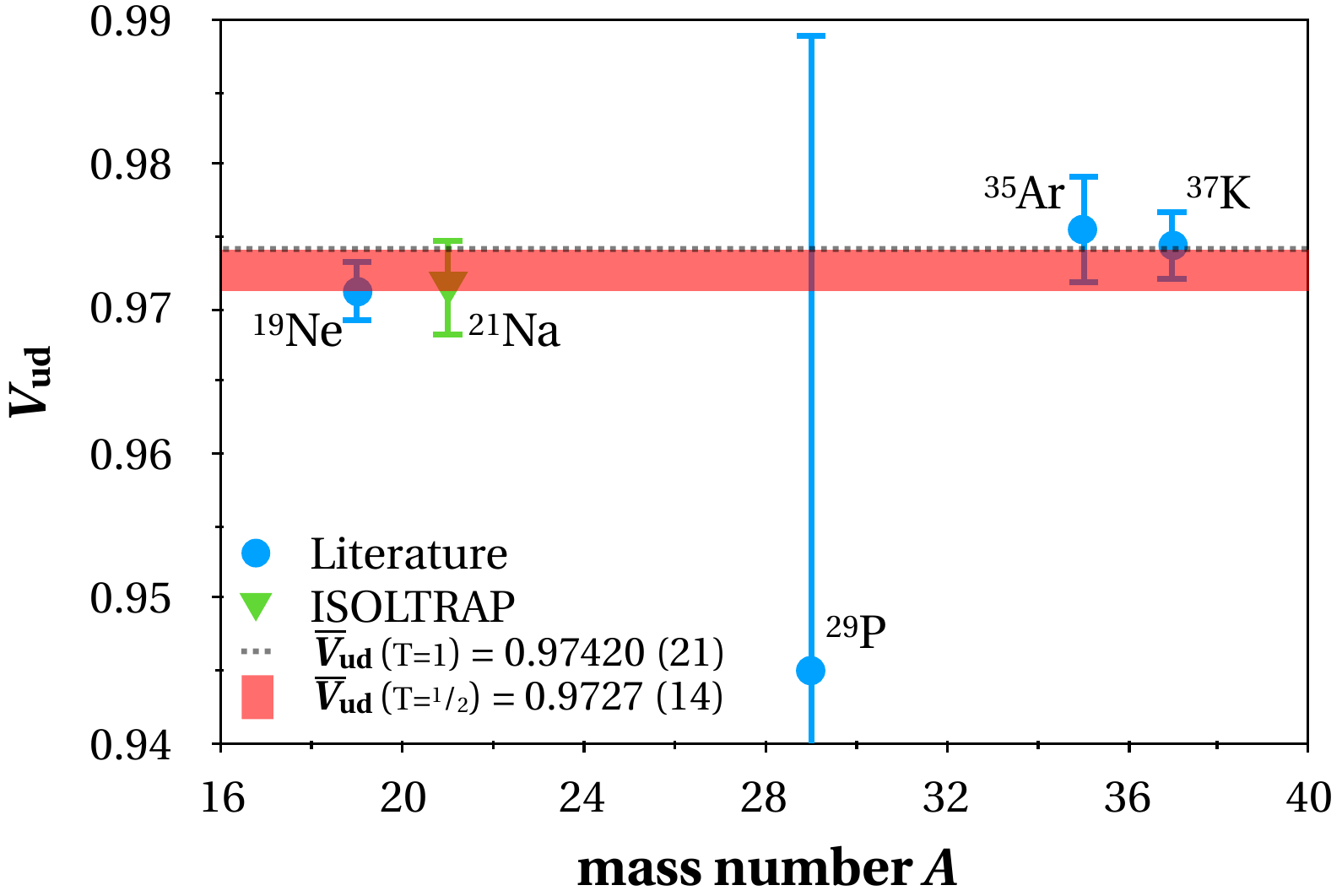}
    \caption{Comparison between different mirror-nuclei $V_{\textrm{ud}}$-values: $^{19}$Ne \cite{Broussard2014}, $^{29}$P \cite{Eibach2015}, $^{35}$Ar \cite{Naviliat-Cuncic2009}, $^{37}$K \cite{Fenker2018} (blue), our new value for $^{21}$Na (green), the weighted mean $V_{\textrm{ud}}$-value of all values (red = two $\sigma$ band) and the mean $V_{\textrm{ud}}$-value for the superallowed transitions \cite{Hardy2018} (gray = two $\sigma$ band).}
    \label{fig:Vud-comparison}
\end{figure}

\noindent The $V_{\textrm{ud}}$-element of the CKM-Matrix
\begin{equation}
    V_{\textrm{ud}} = \sqrt{ \frac{K}{\mathscr{F} t^{\textrm{mirror}}_0 \, G^2_F \, C^2_V \, (1+\Delta^V_R) } }
\end{equation}

\noindent can finally be calculated using $K/(\hbar c)^6 = 2\pi^3ln(2)\hbar(m_ec^2)^{-5} = 8120.276(5) \times 10^{−10}\,$GeV$^{−4} $s, the fundamental weak interaction coupling constant $G_F/(\hbar c)^3 = 1.16639(1) \times 10^{−5}\,$GeV$^{−2}$, the conserved vector current (CVC) constant $C_V=1$ (assuming that the CVC hypothesis is correct, see e.g. \cite{Hardy2015}) and the transition independent correction $\Delta^V_R = 0.02361(38)$\textsuperscript{\ref{footnote1}} \cite{Severijns2008}. The results are summarized in Tab.$\>$\ref{tab:calculations}. In Fig.$\>$\ref{fig:Vud-comparison} a comparison with five other transitions for which a $V_{\textrm{ud}}$ value can be experimentally determined is presented. A comparison between the average $V_{\textrm{ud}}$-value extracted from these mixed Fermi/Gamow-Teller transitions and that extracted using the superallowed transitions \cite{Hardy2018} is also shown in Fig.$\>$\ref{fig:Vud-comparison}.\\

\begin{table}
\caption{\label{tab:calculations}Calculated vector part of the statistical-rate function $f_V$, mirror-nuclei $\mathscr{F} t^{\textrm{mirror}}$-value and $V_{\textrm{ud}}$-element of the CKM-Matrix for $^{21}$Na and $^{23}$Mg. For details, see text.}
\begin{ruledtabular}
\begin{tabular}{lllll}
Isotope  & $f_{\textrm{V}}$ & $f_{\textrm{A}}/f_{\textrm{V}}$ & $\mathscr{F} t^{\textrm{mirror}}$ (s) & $V_{\textrm{ud}}$     \\
\hline
$^{21}$Na  & 170.714(6) & 1.0170(17) & 4071(4) & 0.9715(34)   \\
\cdashline{1-5}
$^{23}$Mg  &  378.51(2) & 1.0195(20) & 4724(14) & N/A   \\
\end{tabular}
\end{ruledtabular}
\end{table}

\begin{figure*}[t!]
\centering
    \includegraphics[width=\linewidth]{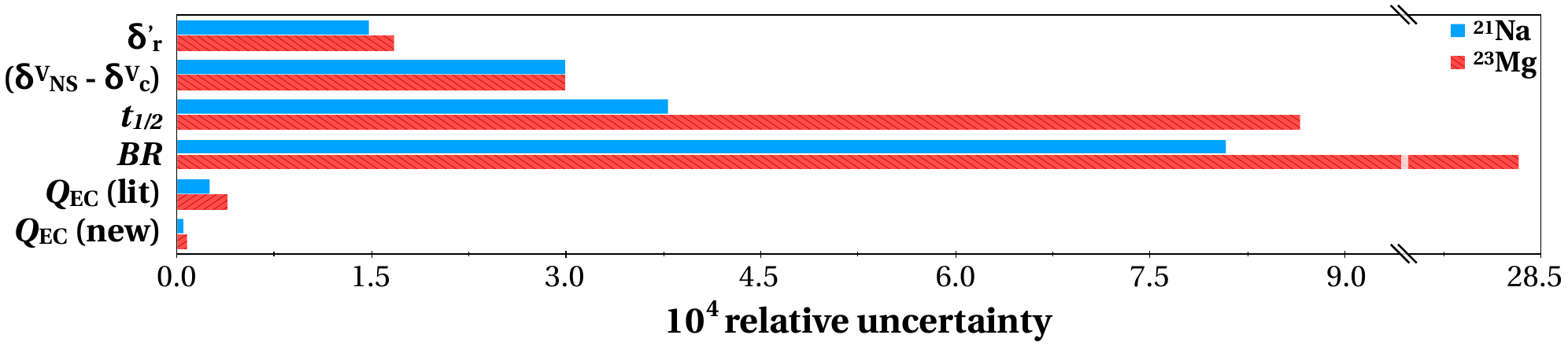}
    \caption{Comparison of the relative uncertainty contributions to the $\mathscr{F} t^{\textrm{mirror}}$-value: the nucleus-dependent radiative corrections, $\delta^{'}_R$, and $\delta^V_{\textrm{NS}}$, the isospin-symmetry breaking corrections $\delta^V_{\textrm{C}}$, the half-lives, the branching ratios $BR$, the $Q_{\textrm{EC}}$ values from \cite{Eibach2015} for $^{21}$Na (blue), from \cite{Schultz2014} for $^{23}$Mg (red), as well as this work. For details, see text.}
    \label{fig:uncertainty-comparison}
\end{figure*}

\noindent The weighted mean of the $V_{\textrm{ud}}$ values for all displayed mirror-nuclei transitions results to $\overline{V}_{\textrm{ud}} = 0.9727(14)$ which is about 7 times less precise than the $\overline{V}_{\textrm{ud}} = 0.97420(21)$ \cite{Hardy2018} of the superallowed transitions. Even though we improved the precision on the $Q_{\textrm{EC}}$-value value for the $^{21}$Na$\rightarrow^{21}$Ne transition by a factor 5, we did not significantly improve the uncertainty on the $V_{\textrm{ud}}$ value for this transition.\\

\noindent Figure \ref{fig:uncertainty-comparison} presents the relative uncertainties attributed to each experimental and theoretical input factor that contributes to the final $\mathscr{F} t^{\textrm{mirror}}$-values. Our new measurements of the $^{21}$Na and $^{23}$Mg $Q_{\textrm{EC}}$-values have such a small relative uncertainty that they do not contribute much to a reduction of the final uncertainty of the $\mathscr{F} t^{\textrm{mirror}}$ values and therefore of the $V_{\textrm{ud}}$ value that can be extracted for $^{21}$Na. As a result our measurements reinforce the motivation for the other experimental quantities, in particular the branching ratios $BR$ and the half-lives $t_{1/2}$, to be measured with significantly improved precision. Furthermore, in case of $^{23}$Mg a $\beta$-asymmetry or $\beta$-neutrino correlation measurement would allow the calculation of an additional mirror-nuclei $V_{\textrm{ud}}$-value.

% ==========================
% Zusammenfassung
% ==========================

\section{Conclusion and outlook}

\noindent This publication presented high-precision $Q_{\textrm{EC}}$-values of the $^{21}$Na$\rightarrow^{21}$Ne and $^{23}$Mg$\rightarrow^{23}$Na mirror $\beta$-transitions with ISOLTRAP at ISOLDE/CERN. Precisions of $\delta m/m = 9 \times 10^{-10}$ and $\delta m/m = 1.5 \times 10^{-9}$ were reached for the masses of $^{21}$Na and $^{23}$Mg, respectively. We reduced the uncertainty of the $Q_{\textrm{EC}}$ values by a factor five, making them the most precise experimental input data for the calculation of the corrected $\mathscr{F} t$-value of these mixed Fermi/Gamow-Teller transitions and strongly reinforces the motivation for improved measurements of the branching ratios $BR$ and the half-lives $t_{1/2}$. Yet lower uncertainties on $Q_{\textrm{EC}}$-values are now reachable with the recently implemented phase-imaging ion-cyclotron-resonance technique \cite{Eliseev2014} which has already been applied to the case of $^{163}$Ho$\rightarrow^{163}$Dy \cite{Eliseev2015}.

% ==========================r
% Danksagung
% ==========================

\section{Acknowledgments}

We thank the ISOLDE technical group and the ISOLDE Collaboration for their professional assistance. We acknowledge support by the Max Planck Society, the German Federal Ministry of Education and Research (BMBF) (05P12HGCI1, 05P12HGFNE, and 05P15ODCIA), the French IN2P3, the Flemish FWO, and the European Union’s Horizon 2020 research and innovation programme (654002). Jonas Karthein acknowledges support by a Wolfgang Gentner Ph.D. scholarship of the BMBF (05E12CHA).

\end{document}